\documentclass[english,aps,pra,reprint,floatfix,showpacs]{revtex4-1}
\usepackage{ifdraft}
\usepackage{babel}
\usepackage[rgb]{xcolor}
\usepackage[final]{graphicx}
\usepackage{amsmath,amssymb}
\usepackage[final]{hyperref}
\usepackage[all]{hypcap}
\newcommand{\be}{\begin{equation}}
\newcommand{\ee}{\end{equation}}
\newcommand{\bea}{\begin{eqnarray}}
\newcommand{\eea}{\end{eqnarray}}
\newcommand{\ba}{\begin{array}}
\newcommand{\ea}{\end{array}}
\newcommand{\ben}{\begin{enumerate}}
\newcommand{\een}{\end{enumerate}}
\newcommand{\bei}{\begin{itemize}}
\newcommand{\eei}{\end{itemize}}

\newcommand{\om}{\omega}

\newcommand{\s}{\sigma}

\newcommand{\la}{\langle}
\newcommand{\ra}{\rangle}

\newcommand{\ad}{a^{\dagger}}
\begin{document}
\author{E. Ginossar}
\author{Lev S. Bishop}
\author{D. I. Schuster}
\author{S. M. Girvin}
\affiliation{Departments of Physics and Applied Physics, Yale University, New Haven, Connecticut 06520, USA}
\title{Protocol for high fidelity readout in the photon blockade regime of circuit QED}
\begin{abstract}
The driven-damped Jaynes-Cummings model in the regime of strong coupling is found to exhibit a coexistence between the quantum photon blockaded state and a quasi-coherent bright state. We characterize the slow time scales and the basin of attraction of these metastable states using full quantum simulations. This form of bistability can be useful for implementing a qubit readout scheme that does not require additional circuit elements. We propose a coherent control sequence that makes use of a simple linear chirp of drive amplitude and frequency as well as qubit frequency. By optimizing the parameters of the system and the control pulse we demonstrate theoretically very high readout fidelities ($>98\%$) and high contrast, with experimentally realistic parameters for qubits implemented in the circuit QED architecture.
\end{abstract}

\maketitle

Qubit readout in solid state systems is an open problem, which is currently the subject of intensive experimental and theoretical research.
High fidelity single shot readout is an important component for the successful implementation of quantum information protocols, such as measurement based error correction codes \cite{nielsen_chuang} as well as for closing the measurement loophole in Bell tests \cite{Garg_Mermin_Bell_1987,*kofman_analysis_2008, ansmann_violation_2009}. For measurements where the observed pointer state depends linearly on the qubit state, for example dispersive readout in circuit QED (cQED) \cite{blais_cavity_2004}, there exists a unified theoretical understanding \cite{clerk_rmp_2009}. Experimentally, these schemes require a following amplifier of high gain and low noise, spurring the development of quantum limited amplifiers \cite{Lafe_amplifier,*bergeal-2009, *castellanos-beltran_amplification_2008}. However, the highest demonstrated fidelities to date rely on nonlinear measurement schemes with qubit dependent latching into a clearly distinguishable state, e.g. Josephson Bifurcation Amplifier (JBA) as well as optimized readout of phase qubits \cite{siddiqi_dispersive_2006, *mallet_single-shot_2009, ansmann_violation_2009}. For this class, during the measurement, the system evolves under the influence of time varying external fields and nonlinear dynamics, ultimately projecting the qubit state. The space for design and control parameters is very large, and the dependence of the readout fidelity on them is highly nontrivial. Therefore the optimization is difficult and does not posses a generic structure.

\begin{figure}[h]
\includegraphics{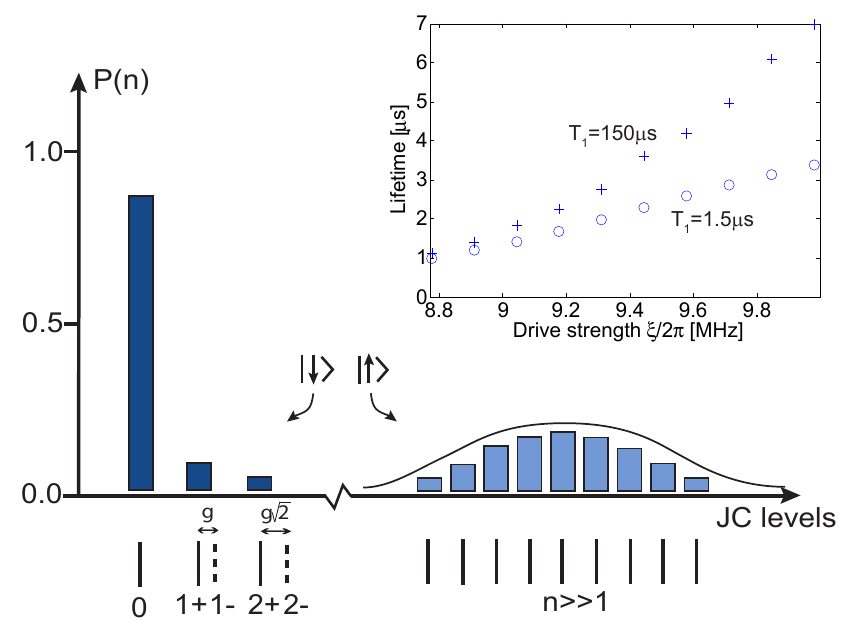}
\caption{\label{intro_figure} A schematic of the dim state (left histogram) in the quantum part of the Jaynes-Cummings (+) manifold  and the QCS in the semi-classical part (right histogram). Application of a readout pulse sequence dynamically maps the initial qubit state into one of the states with high fidelity. The inset shows the lifetimes of several QCS with increasing drive strength and mean photon number, while for these parameters the dim state is photon blockaded (the coexistence regime), and the qubit decay ($T_1$) has a distinct influence on the lifetime ($\tau_b$) of the QCS.}
\end{figure}

\begin{figure}
\includegraphics{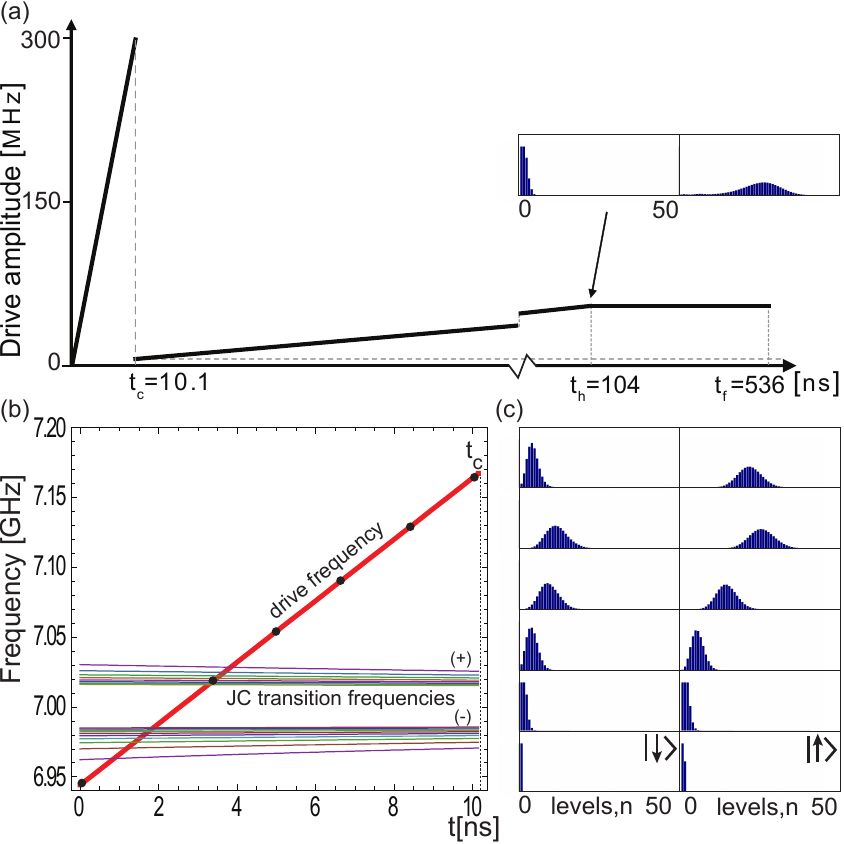}
\caption{\label{chirp_figure}Readout control pulse (a) Time trace of the drive amplitude: a fast and selective initial chirp (10ns) can selectively steer the initial state, while the qubit is detuned from the cavity ($(\om_q-\om_c)/2\pi \approx 2g$). It is followed by a slow displacement to increase contrast and lifetime of the latching state, while the qubit is resonant with the cavity ($\kappa/2\pi=2.5 \textrm{MHz}$). The drive amplitude ramp is limited so that the photon blockade is not broken, but the contrast is enhanced by additional driving at the highest drive amplitude. (b) A diagram of transition frequencies shows how the drive frequency chirps through the JC ladder frequencies of the (+) manifold, and how the manifold changes due to the time dependent qubit frequency. (c) Wave packet snapshots at selected times (indicated by bullet points on panel (b)) of the chirping drive frequency of panel (b) conditioned on the initial state of the qubit.}
\end{figure}

\begin{figure}
\includegraphics{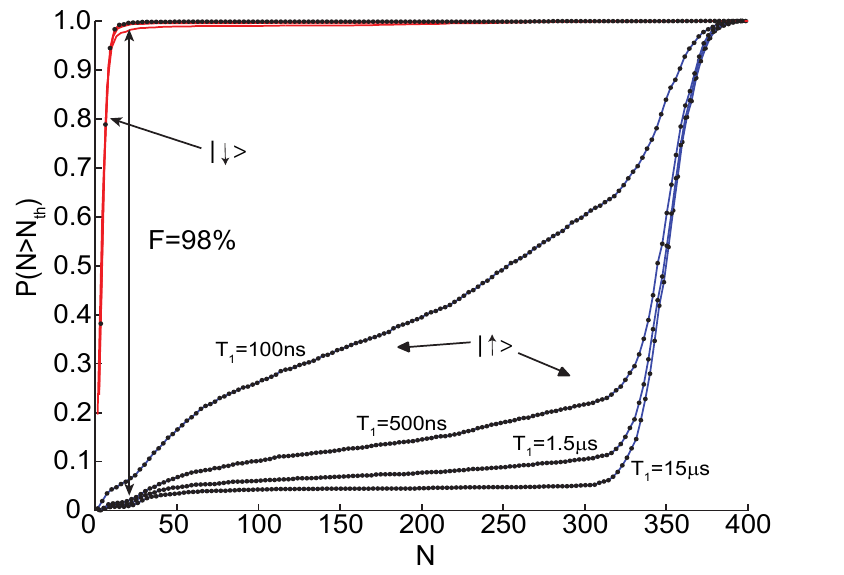}
\caption{\label{scurve_figure}Cumulative probability distributions for the number of photons (N) emitted from the cavity during the driving time $t_f$, for different qubit decay times ($T_1$), including a very long $T_1=15\mu s$ indicating that $T_1$ is not limiting the readout fidelity. For low detection thresholds ($N_{th}\approx 20$) for distinguishing $|\hspace{-0.05in}\uparrow\ra\,(N>N_{th})$ from $|\hspace{-0.05in}\downarrow\ra \,(N<N_{th})$ the fidelity can be very high ($>98\%$) for realistic values of qubit decay in cQED (a few microseconds), and of high contrast for more moderate fidelities ($>90\%$). The distributions also show a almost no false positives for higher thresholds $N_{th}>20$ (here $t_h=194ns$ and other parameters as for Fig.~\ref{chirp_figure}).}
\end{figure}

We propose a coherent control based approach to the readout of a qubit that is strongly coupled to a cavity, based on an existing cQED architecture, but not necessarily limited to it. This approach is in the spirit of the latching readout schemes, but it differs in that the source of the nonlinearity is the Jaynes-Cummings (JC) interaction. When the qubit is brought into resonance with the cavity mode, the strong anharmonicity of the JC ladder of dressed states can prevent the excitation of the system even in the presence of a strong drive, a quantum phenomenon known as photon blockade \cite{imamolu_strongly_1997,*birnbaum_photon_2005, *houck-aps-2010}. However, due to fact that the JC anharmonicity is diminishing with the excitation number, we find a form of bistability, where highly excited quasi-coherent states (QCS) co-exist with the blockaded dim states (Fig.~\ref{intro_figure}).
In order to make use of this bistability to read out the qubit, it is necessary to solve the coherent control problem of
selective population transfer, which is how to steer the system towards either the dim state or the QCS, depending on the initial state of the qubit (Fig.~\ref{chirp_figure}). This selective dynamical mapping of the qubit state to the dim/bright states constitutes the readout scheme. It is potentially of high contrast, and hence robust against external amplifier imperfections.
An advantage of this readout is that it uses no additional components beyond the qubit and the cavity, the latter already present as part of the cQED architecture. Based on a full quantum simulation which includes dissipation of the qubit and the cavity (we ignore pure dephasing \footnote{Typically transmon qubits operate in a regime of $E_J/E_C\gg 1$ where pure dephasing is exponentially suppressed.}), we predict that implementing this scheme should yield very high fidelities between  $90\%$ and $98\%$ for a typical range of realistic cQED parameters (Fig.~\ref{scurve_figure}).

We consider the Jaynes-Cummings model with drive and dissipation ($\hbar=1$)
\be H=\om_c \ad a + \frac{\om_q}{2} \s_z + g(\s_+ a + \ad \s_- ) + \xi(t) (a+\ad) + H_{\gamma} + H_{\kappa} \ee
where $\xi(t)$ is the time-dependent drive of the cavity, $g$ is the cavity-qubit coupling, and $H_{\gamma,\kappa}$ represent the coupling to the qubit and cavity baths, respectively.
When the system is initialized in the ground state, there is a range of drive strengths for which the system will remain blockaded from excitations out of the ground state. However, since the anharmonicity of the JC ladder decreases with excitation number, the transition frequency for excitations between adjacent levels ultimately approaches the bare cavity frequency. Qualitatively, when the excitation level $n$ is such that the anharmonicity becomes smaller than the linewidth $\kappa$, we expect the state dynamics to be semiclassical, similar to a driven-damped harmonic oscillator \cite{alsing_spontaneous_1991,*kilin_single-atom_1991}. More specifically, in order to support a coherent wave packet centered around level $n$, with a standard deviation of $\sqrt{n}$, the difference of transition frequencies across the wavepacket has to be of the order of the linewidth $\kappa$. This approximate criterion for a minimal $n$ can be written as $\omega_{n+2\s}-\omega_{n-2\s}\approx \kappa$
where $\om_{n\pm 2\s}$ are the ladder transition frequencies, positioned $2\s=2\sqrt{n}$ above and below the mean level $n$. Quantitatively, we find in simulations that the lifetime of the QCS is long but finite, and increases with the amplitude of the drive (see insert in Fig.~\ref{intro_figure}). As we explain below, we find that low lying QCS states ($\bar{n}=20$) are the most effective for optimizing the overall readout fidelity. Note that the JC ladder consists of two manifolds (originating from the degeneracy of the bare states $|g, n+1\ra$ and $|e, n\ra$) denoted by $(\pm)$, and we will always refer to a states occupying one manifold since the drive is off-resonant with respect to the other manifold. Transitions between manifolds contribute to the decay of the QCS to the dim state. Such transitions can be induced by the drive but their rate is smaller by a factor of $\mathcal{O}(n^{-1/2})$ compared to the rate of transitions inside the same manifold. An additional source of inter-manifold transitions are decay ($T_1$) and pure qubit dephasing $(T_{\varphi})$, whose effects in the presence of drive were studied in the context of the dispersive regime \cite{boissonneault_dispersive_2009}. Indeed, as we see in Fig.~\ref{intro_figure}, changing $T_1$ has a noticeable effect on the QCS lifetimes. For very large $\bar{n}$, these processes become ineffective for inducing decay of QCS, since then the difference between manifolds excitation frequencies becomes smaller than $\kappa$, and therefore the drive effectively drives both manifolds. For superconducting transmon qubits $T_1$ is the dominant decay process, and we show its effect on the overall fidelity in Fig.~\ref{scurve_figure}.\looseness=-1
\begin{figure}
\includegraphics{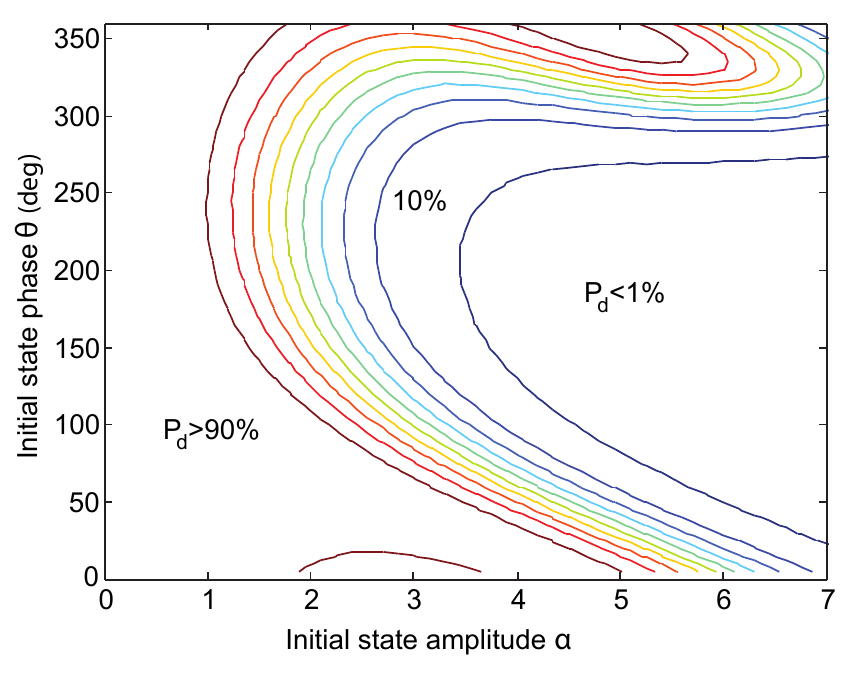}
\caption{\label{basin_figure}Probability for the QCS to decay after being initialized as coherent state wave packet $|\alpha e^{i\theta}\ra$ and driven for a time $\kappa^{-1}$. The equal probability contours trace out two basins of attraction: states initialized inside the low decay probability contour ($P_d<1\%$) end up long lived ($\tau_b\gg \kappa^{-1}$), whereas states initialized left of $P_d=90\%$ quickly decay to the ground state and remain photon blockaded. The parameters are $g/2\pi=100\textrm{MHz}$, $\kappa/2\pi=4.05\textrm{MHz}$, $\xi/2\pi=9.9\textrm{MHz}$,$(\om_{d}-\om_{c})/2\pi=12.3\textrm{MHz}$,$(\om_{c}-\om_{q})/2\pi=9.7\textrm{MHz}$, $T_1=1591ns$ ($\om_d$ is the drive frequency).}
\end{figure}

The QCS exist with drive amplitudes where the ground state is photon blockaded, giving
rise to a dynamical bistability between quantum and semi-classical parts of the JC ladder. Indeed, we see that there is a basin of attraction for states initialized as coherent wave packets to persist as QCS, and we characterize it according to the probability of the state to decay on the timescale $\tau_b \gg \kappa^{-1}$. In Fig.~\ref{basin_figure} we plot the contours of equal probability of the QCS to decay to a manifold of states close to the ground states, after a time $\kappa^{-1}$, given that it was initialized with a certain amplitude ($\alpha$) and phase ($\theta$). We see a large region supporting QCS, and the phase sensitivity can be understood qualitatively from the time dependent simulations: a mismatch between the phases of the drive and initial coherent state causes ringing of the wave packet outside of the basin into the too anharmonic part of the ladder from which it cannot recover. In addition to the existence of this basin, the anharmonicity acts together with the cavity decay to induce mixing of the QCS: even for bright states ($\bar{n}>40$) we extracted a relatively low purity of $Tr(\rho^2)<0.5$.

In the regime of coexistence the dim quantum state and QCS present us with the possibility of implementing a high contrast readout scheme. This requires the solution of the coherent control problem of steering the logical $|\hspace{-0.05in}\uparrow\ra$ state to some point on within the basin of attraction (Fig.~\ref{basin_figure}), while keeping the $|\hspace{-0.05in}\downarrow\ra$ far from the basin, in the manifold of dim states. In the presence of dissipation, the latter would quickly decay to the ground state, and remain there even in the presence of driving, due to the photon-blockade, whereas
the QCS would persist for a long time $\tau_{b}$ and emit approximately $\kappa \la n \ra_{b} \tau_{b}$ photons.
The standard coherent control problem of population transfer \cite{bergmann_coherent_1998}, which was also discussed recently for superconducting qubits \cite{Jirari-EPL}, is to maximize the probability $P_{i\rightarrow f}$ of steering the state $|i\ra$ to the state $|f\ra$. However, here the goal is to bring the probability for {\em selective} steering $P_{i\rightarrow f}+P_{i'\rightarrow f'}<2$ close to its theoretical maximum, which is an essentially different coherent control problem. For systems with very large anharmonicities, for example atomic systems it is possible to effectively implement a population transfer via adiabatic control schemes such as STIRAP \cite{bergmann_coherent_1998}. The JC ladder anharmonicity is relatively small compared to atomic systems, and so these schemes are inapplicable here.

The control pulse sequence we apply is depicted in Fig.~\ref{chirp_figure}, and consists of three parts: (1) a strong chirped pulse ($t<t_c$) drives the cavity, with the qubit being detuned such that the cavity frequency is state dependent. This pulse maps the $|\hspace{-0.05in}\downarrow\ra$ and $|\hspace{-0.05in}\uparrow\ra$ to the dim and bright state basins, respectively. (2) a weaker pulse that transfers the initially created bright state to even brighter and longer lived states ($t_c<t<t_h$), and (3) steady driving for additional contrast ($t_h<t<t_f$). In designing such a pulse sequence we have the following physical considerations: (a) the initial fast selective chirp has to be optimally matched to the level structure so that the population transfer and selectivity would be extremely high (b) it is necessary to chirp up quickly before decay processes become effective and result in false negative counts ($t_c\kappa\approx 0.16$) (c) for $t>t_c$ it is necessary to drastically reduce drive strength, since it reaches drive strengths which would break the photon blockade through multiphoton processes if it persisted. The piecewise linear chirp sequence is fed into a full quantum simulation that includes decay, and the 13 parameters of the system and drive are optimized with respect to the total readout fidelity.

The cumulative probability distributions to emit $N$ photons conditioned on starting in two initial qubit states are plotted in Fig.~\ref{scurve_figure}. These distributions were optimized for $T_1=1\mu s$ and then regenerated after varying $T_1$, in order to depict the effect of qubit relaxation on the readout. There are two figures of merits from figure \ref{scurve_figure}: one is that there exist very high fidelities $\mathcal{F}=1-P(\uparrow|\downarrow)-P(\downarrow|\uparrow)$ exceeding $98\%$ for a low threshold around $N_{th}=20$ even for relatively short lived qubits ($T_1\approx 500ns$). In order to take advantage of these fidelities a very low noise amplifier would be needed.
In addition, we find high contrast and high fidelities ($>90\%$) for long lived qubits $T_1>1.5\mu s$ with thresholds around $N_{th}=150$. These fidelities will not be limited by the noise added by a cryogenic HEMT amplifier with noise temperature of $T_N\approx 5K$. For $T_N>5K$ the effect of the noise can be mitigated by increasing $t_f$ (up to time $\tau_b$). Note also that the limit of obtainable fidelities with this control scheme is not due to finite qubit lifetime, as we see from the curve that was simulated for $T_1=15\mu s$. Another useful feature of these distributions is the very low level of false positives (red curve for qubit state $|\hspace{-0.05in}\downarrow\ra$), for a wide range of thresholds, originating from the effectiveness of the photon blockade.

For experimental applications it is also important to know how robust the fidelity is against deviations of the control pulse parameters from their optimal values. We therefore varied the parameters of the initial chirp pulse $\delta_{d,c}=\om_d-\om_c,\dot{\delta}_{d,c}, \delta_{d,q}=\om_d-\om_q, \dot{\delta}_{d,q}, \dot{\xi}, t_c,$ and $\kappa$ independently around their optimal values and checked the range for which the fidelity is above $98\%$. The fidelity is most sensitive to variations of the duration of the chirp pulse $t_c$ which yields a tolerance of $\pm 10\%$ and higher ranges ($>\pm 20\%$) for the rest of the parameters.

For cQED all the necessary components for the above scheme have been experimentally demonstrated. Strong qubit-cavity coupling has been demonstrated in many experiments \cite{thompson_observation_1992,raimond_colloquium:_2001,wallra_strong_2004}. Strong driving of a cavity-qubit system has been shown in \cite{baur_measurement_2009}, with the system behaving in a predictable way, as well as photon blockade \cite{bishop_nonlinear_2009} and fast dynamical control of the qubit frequency via flux bias lines \cite{dicarlo_demonstration_2009}. In addition there is evidence both theoretically and experimentally for the increasing role that quantum coherent control plays in the optimization of these systems for tasks of quantum information processing. As examples we can mention improving single qubit gates \cite{motzoi_simple_2009}, two-qubit gates \cite{fisher_optimal_2009}, and population transfer for phase qubits \cite{Jirari-EPL}. We therefore believe that the readout scheme would be applicable for the transmon, although the control parameters would have to be re-optimized due to the effect of additional levels.

The suggested readout scheme is different from other existing schemes in several aspects. Compared to dispersive readout \cite{blais_cavity_2004} it involves very nonlinear dynamics and could potentially exhibit much higher fidelity and contrast. Even though it relies on a dynamical bistability, it is essentially different from the JBA and another recently proposed scheme \cite{bishop_jco}, since it explicitly operates using the quantum photon blockade. It is important to stress that the above optimization of the control parameters is only partial, since we have limited ourselves to simple linear chirps in this work. More complex modulations are certainly possible although the standard methods for optimal control \cite{khaneja_optimal_2005} may be difficult to implement here due to the large Hilbert space. Therefore we believe that an experimentally based optimization using adaptive feedback control \cite{judson_teaching_1992} might be best option, and has the potential to yield superior readout fidelities for higher detection thresholds.

We acknowledge useful discussions with M.~H. Devoret.
This work was supported by the NSF under Grants Nos. DMR-0603369 and DMR-0653377, LPS/NSA under ARO Contract No. W911NF-09-1-0514, and in part by the facilities and staff of the Yale University Faculty of Arts and Sciences High Performance Computing Center.

\bibliography{jcbib,Lev_Thesis}
\end{document}